\documentclass{llncs}

\usepackage[pdftex]{xcolor}
\usepackage[hyphens]{url}
\usepackage{listings}
\usepackage{xspace}
\usepackage{graphicx}

\begin{document}

\title{Twenty-Five Shades of Greycite: Semantics for referencing and preservation}
\author{Phillip Lord, Lindsay Marshall}
\institute{School of Computing Science, Newcastle University}
\maketitle

\begin{abstract}
  Semantic publishing can enable richer documents with clearer,
  computationally interpretable properties. For this vision to become reality,
  however, authors must benefit from this process, so that they are
  incentivised to add these semantics. Moreover, the publication process that
  generates final content must allow and enable this semantic content. Here we
  focus on author-led or ``grey'' literature, which uses a convenient and
  simple publication pipeline. We describe how we have used metadata in
  articles to enable richer referencing of these articles and how we have
  customised the addition of these semantics to articles. Finally, we describe
  how we use the same semantics to aid in digital preservation and
  non-repudiability of research articles.
\end{abstract}

\section{Introduction}
\label{sec:introduction}

The academic publishing industry is changing rapidly, partly as a result of
external changes such as the move to open access, and partly as a final
recognition in the importance of the web. With change comes the opportunity to
add more semantics to
publications~\cite{shadbolt2006semantic,shotton2009semantic,shotton2009adventures},
to increase the computational component of papers, enabling publication to
take its place in the linked data environment~\cite{bizer2009linked}.

While within academia, third party publication --- where knowledge is given to
a third party to manage the publication process --- is commonplace, outside in
many technical disciplines we see direct publication, where the author
publishes work that readers can then directly access. This form of publication
is often called ``grey literature'' publication --- a somewhat derogatory term ---
however, it has some significant advantages. It is rapid and places the author
in control, allowing them to innovate in terms of presentation and
content\cite{procter12:_use_twitt_faceb_phd_studen_schol_commun}. It operates
without editorial control from third-party publishing which may help to
overcome the publication bias found in many areas of scientific publishing. We
have previously used a form of grey literature publishing to publish ontology
tutorial material\cite{greycite2127,greycite1341}; this has resulted in the
release of useful material which would otherwise probably not have been
created, as many academics regard book chapters as having little
purpose\cite{greycite8303}.

From the perspective of semantic publishing, it has an additional advantage;
the process is often very simple, without additional human intervention
between the author and the final published form. This simplicity means that
semantics added by the author can pass through to the published version with
relative ease. In the process, it is also possible that semantics added by the
author can aid in the authoring process, which we consider of
key importance\cite{greycite1325}.

However, grey literature publishing lacks some of the formality of third-party
academic publishing; for instance, several organisations provide centralised
collection of bibliographic metadata; we have used this metadata, for
instance, to enable accurate citation of academic literature through the use
of primary identifiers. The lack of a centralised authority for author
published literature, however, prevents this technique from being used for
general URIs. This presents us with a simple research question: are there enough
semantics on the extant web to provide clear bibliographic metadata for
different web pages?

In this paper, we describe two new systems: greycite and kblog-metadata. The
former, addresses the problem of bibliographic metadata, without resorting to
a single central authority, extracting this metadata directly from URI
endpoints. The latter provides more specialised support for generating
appropriate metadata. We describe how these systems support our three steps
doctrine~\cite{greycite1325}, which suggests that semantic metadata must be of
value to all participants in the publishing process including the authors. We
also describe how these systems can impact on another major problem with
author-led publishing: that of archiving and ``link-rot''.

\section{References}
\label{sec:referencing}

Referencing is ubiquitous within scientific and academic literature, to the
extent that it can be considered to be a defining feature. Academics reference
previous work both as a utility to the reader, and as a mechanism for
establishing provenance. However, reference insertion and formatting is
complex to the point of humour\cite{greycite8297}; with nearly 3000 citation
formats in common use\cite{greycite8305}, reversing the process is even
harder.

We have previously described the \emph{kcite} tool which enables automatic
generation of reference lists from a primary identifiers\cite{greycite1325}:
as described previously, it is often possible to hide these from the user
behind tooling, so that they do not need to insert primary identifiers by
hand\cite{greycite1325}. This form of referencing also has advantages for
human and machine consumpution of the data; the primary identifier, which is
also accessible to downstream analysis; moreover, because the reference is
generated as a result of this identifier, when the author checks the
reference, they are also effectively checking the identifier, which
conventionally, the author must check manually at extra cost to their time.As an
\textit{ad hoc} measure, user feedback from our tool has now identified a
number of primary identifiers (DOIs) with inaccurate metadata, and one
systematic error in the presentation of these identifiers affecting many
institutional repositories\cite{greycite8865}.

This, however, requires a source of metadata: currently, kcite supports (most)
DOIs, arXiv and PubMed IDs directly, all of which allow metadata harvesting.
Following the development of kcite, our request resulted in both CrossRef and
DataCite -- the two most significant DOI registration agencies for academia --
providing metadata in the form that kcite consumes (Citeproc JSON). For
general URIs, unfortunately, there is no centralised authority which can
provide this metadata. 

\subsection{Technical Glossary}

Here, we provide a short technical glossary of the tools described, also shown
in Figure~\ref{fig:architecture}, as an aid to understanding.
\begin{description}
\item[kcite:] A wordpress plugin that generates a reference list for an article
  from primary identifiers. Uses a variety of services, including greycite, to
  resolve identifiers to bibliographic metadata.
\item[kblog-metadata:] A wordpress plugin that provides flexible presentation
  of bibliographic metadata, both computationally and visibly through
  on-screen widgets.
\item[greycite:] A server which returns bibliographic metadata for any URI,
  extracted from that URI for the article resolved by that URI.
\item[Citeproc JSON:] A bibliographic format defined by the Citeproc-js tool.
\item[BibTeX:] a format defined by the BibTeX tool.
\end{description}

\section{The Greycite System}
\label{sec:techn-descr}

We initially considered the possibility that kcite could harvest its own
metadata directly. It would have been possible, for instance, for a kcite
installation on one site to return metadata to another, through a REST call,
or as embedded metadata. However, this would have required users to know in
advance which URIs were so enabled, and would have worked with few websites.

To avoid this limitation, we wished to use extant semantics already on the
web; the complexity of this task argued against integration with kcite which
is an end-user tool. Additionally, as a server greycite would usable to more
than one client; in fact, this has proven to be the case, with a third-party
tool, knitcitations which supports dynamic citations in a literate programming
environment for R\cite{greycite8222}.

\begin{figure}
  \centering
  \includegraphics[width=0.7\textwidth]{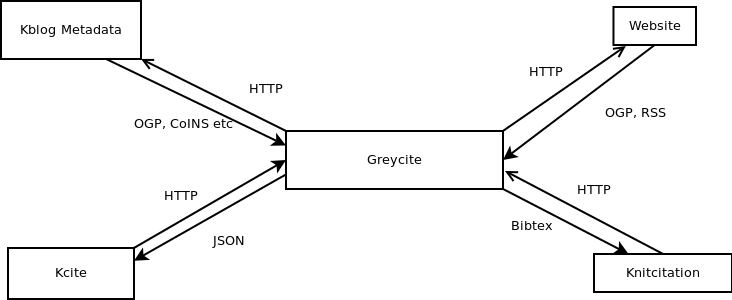}
  \caption{Client server interaction between Greycite and clients}
  \label{fig:architecture}
\end{figure}

Greycite provides bibliographic metadata in a variety of formats on request
about an arbitrary URI; an architectural overview is shown in
Figure~\ref{fig:architecture}. It uses a simple REST API to do this, and
returns either Citeproc-JS JSON (directly used by kcite)\cite{Citeproc-js},
BibTeX (used by knitcitations, and the kblog-metadata tool described here). We
have additional support for other formats, including RDF (encoding Dublin
Core), RIS, and Wikipedia ``cite'' markup. It store the results of metadata
extractions, initially for reasons of efficiency, although this is also
valuable for ephemeral sources of metadata(see
Section~\ref{sec:sources-metadata}).

Greycite extracts a number of sources of metadata, and uses a scoring scheme
and a set of heuristics to choose between them; we describe these next. 

\section{On how the Web describes itself}
\label{sec:sources-metadata}

To enable referencing, we need five key pieces of bibliographic metadata, namely:
\begin{itemize}
\item Author(s) (A)
\item Title (T)
\item Date of publication (D)
\item ``Container'' -- equivalent to journal, conference or website. (C)
\item Canonical Identifier (I)
\end{itemize}
These are the minimal pieces of metadata used by most referencing styles, and
following standard publication practices. We have now investigated many
sources of web-delivered metadata. These have been discovered in a number of
ways: some were designed for this purpose. Others, where discovered by
inspection of academic websites; some were discovered entirely by chance
(where an authors name was visible on a web page, but not extractable, we
search for all instances of that name, looking for structure). We prioritised
``interesting'' websites, for our definition of interesting.

A complete list of all the mechanisms greycite uses for metadata extraction
is shown in Table~\ref{tab:sources}. By itself HTML provides very few of
these five pieces of metadata; only the title is extractable; even
here, for most browsers, the title is displayed publicly, in the browser
title bar. As a result, many sites include the name of the site, often ``|''
delimited in the title of each page, which makes this a relatively messy form
of data.

We also investigated the use of CoINS metadata; this standard is used by a
number of academic websites, and can be consumed by a some bibliographic
tools\footnote{\url{http://ocoins.info}}. It is an imperfect tool. The
standard is rather confusing to read, the main website describes it as using a
NISO 1.0 Context Object, the link to the specification for which is broken.
Different implementations tend to produce different variations of the same
metadata. More over, CoINS metadata does not necessarily describe the article
being posted; for example, \url{http://researchblogging.org} uses CoINS to
describe a secondary article being reviewed. It has a significant advantage,
however, over most metadata specifications which is that it is embeddable in
the \emph{body} of a web page; for hosted websites, authors often do not
control the headers and cannot add elements to it.

The guidelines for inclusion to Google Scholar are somewhat clearer, and
easier to implement, although even here there are common causes for confusion
(\texttt{citation\_author} vs \texttt{citation\_authors}). This form of
metadata is relatively common on many journal websites, but is not, in our
experience, wide-spread outside academia. More common, is Open Graph Protocol,
or OGP\footnote{\url{http://ogp.me}}; this is a form of RDFa developed as part
of the Facebook platform. It is found on a large number of websites including
many common blog platforms, as well as various news outlets, such as BBC News,
which are otherwise hard to cite. The author list is often not represented in
OGP\footnote{Including on the OGP website!}; while OGP has the ability to do
this, authorial metadata needs to be gathered from a secondary URI, linked
from the main content; this is more complex to implement, which may explain
why it is commonly missing.

\begin{table}
  \centering
  \small
  \begin{tabular}{l|l|p{0.8\textwidth}}
Source & Type & Notes \\ 
\hline
Atom & TDCAI & Inferences where article is not present\\   
CoINS & TCDA & Blocked where identifier does not match location\\
CEUR-WS & TCDA & Uses span tags in index files\\
Dublin Core & TCDA & Both \texttt{dc:} and \texttt{dc.} recognised\\
Eprints & TCDA & \\
EXIF & TDA & In Progress\\
FOAF & N/A & In Progress\\
GIF & N/A & In Progress\\
Google Scholar & TCDA & Both \texttt{citation\_author} and \texttt{citation\_authors}. Bepress prefix with \texttt{bepress} \\
HTML & T & The ``title'' tag\\
Link & N/A &  In Progress\\
Meta & TCDA & Common uses recognised\\
OGP & TCDAI & Some syntactic variants\\
OpenLibrary & TCDI & Preliminary\\
ORCID & TCA & Screen Scraping\\
PDF & TA & Often fails!\\
Prism & CD & \\
RSS & TCDAI & See Atom\\
Schema & TD & In Progress\\
Scholarly HTML & N/A & In Progress. Never seen in the wild\\
ScienceDirect & TCD & Screen Scraping\\
Twitter & TCAI & ``Author'' is normally a hashtag\\
URI & D & Heuristic based on link structure\\
W3C & TCDAI & Screen scraping specific for W3C specifications\\
WorldCat & TDA & Screen scraping\\
  \end{tabular}
  \caption{Twenty-Five Sources of Metadata: type indicates the metadata extractable (\textbf{T}ype,
    \textbf{D}ate,\textbf{C}ontainer,\textbf{A}uthor,\textbf{I}dentifier).In
    progress indicates that we believe more metadata is present. Screen
    Scraping means heuristics based on HTML structure.}
  \label{tab:sources}
\end{table}

Another source of authorial metadata is RSS/Atom feeds. Many common platforms
include a \texttt{dc:creator} tag and this is often the only easily
extractable form of metadata. We do find that generic (\texttt{admin},
\texttt{blog admin}) or personal but informal (\texttt{Phil},
\texttt{phillord}) user names are fairly common; this is the default behaviour
for many content management systems, and appears to be a conscious choice for
many multi-user sites. Greycite filters some of the more common ones and does
not consider them as valid metadata. We also provide heuristics where articles
are missing; for instance, if all articles in an RSS feed have the same author
and container title, we infer this information for missing articles.

Another commonly missing piece of metadata is date; while it can be found in
RSS/Atom feeds, these are not always present and are \emph{ephemeral}. In
contrast to author or container information, publication date cannot be
infered where articles are missing from metadata on other articles. We apply a
heuristic here in acknowledgement of the fact that many blogs use a date
format for their URI permalinks. In fact of the URIs in greycite, we can mine
date metadata from some 33\% of them; while this is not a representative
sample, it does show that heuristics can be surprisingly effective.

Unfortunately, many scientific papers are published in PDF; while we do
attempt to extract metadata from these, greycite is currently not very
effective, so most PDFs appear to contain no extractable metadata; we are
investigating more PDF parsers to attempt to address this problem. In some
cases, we have provided heuristics which work around this difficulty: greycite
will provide metadata for PDFs hosted by CEUR-WS; however, we achieve this by
mining metadata from the HTML files which link to the PDF.

A significant number of websites do not provide any specific metadata that we
were able to discern; interesting and surprising cases include most of the W3C
standards, websites for both the International and Extended Semantic Web
Conferences, and the ORCID webpages. We have a significant number of special
purpose extraction plugins; for instance, from a desire to reference W3C
specifications, we have created a single site plugin which uses a highly
\textit{ad hoc} screen-scraping technique. Taken together, of the 4000 URI that
have been submitted, Greycite can extract the main four pieces of metadata
(TCDA) from 62\% of URIs.

\section{On how the web could describe itself}
\label{sec:kblog-metadata}

While Greycite can extract metadata from many different sources, it does
require some support from the content. Unfortunately, for many content
management systems whether this metadata is available or not is dependent on
the local setup; for instance, with WordPress, the presence or absence of many
sources of metadata is theme dependent; the exception to this is data from the
RSS/Atom feeds although even here, the feeds themselves can be disabled at the
theme level\footnote{This would generally be considered to be a broken theme},
or through author choice\footnote{This would generally be considered to be a
  broken author}. 

We have therefore created a plugin for WordPress to address this need; while
the solution is, of course, specific to WordPress, the use cases that we
address are considerably more general. This plugin, \emph{kblog-metadata},
currently adds metadata in three formats: Google Scholar, OGP and CoINS. The
latter is used by and has been tested with Zotero and similar bibliographic
tools, which is the main reason for its inclusion. Facebook provide an
explicit tool for testing OGP, while Google Scholar do not. By default,
kblog-metadata requires no configuration and uses knowledge directly from the
WordPress container, which provides suitable values for the five pieces of
metadata we require (see Section ~\ref{sec:sources-metadata}).

While the author can check that their metadata is appearing correctly, through
the use of greycite, this requires them to use a secondary website.
Alternatively, they can link to their article using kcite, which will then
generate a reference on the basis of the metadata; however, this requires
creating new content, to check old. Following our three steps doctrine, we
wished to make the metadata more useful for the authors. We have, therefore,
added ``Widget'' support, which displays citation information for each page
(or a website as a whole) using the same metadata resolution techniques; this
display both eases the task of checking the metadata, as well as incentivising
the author to do so. The widget also provides a BibTeX download of the
citation. As well as being useful for authors and readers, this has an
additional utility: the BibTeX actually comes from greycite, on the basis of
its metadata extraction. When anything (including robots) access this BibTeX,
Greycite is invoked, and hence becomes aware of the new article.

Although for simple use, the default WordPress data suffices, there are
several uses cases where it does not. Therefore, kblog-metadata provides
authors with the ability to set the metadata independently on an individual
post basis. This fulfils a number of use cases. The most common of these is
for multiple-author posts; WordPress multiple author support is built around
editing rights, rather than authorship. Hence all authors must have WordPress
logins which they otherwise may neither want or need. Kblog-metadata allows
setting authorship lists independently of login rights. Secondly, authors may
also wish to provide an alternative container title. Combined, these two
facilities enable WordPress to operate as an ``preprints'' server. For
example, \url{http://www.russet.org.uk/blog/2054} resolves to the full text of
our paper\cite{greycite1325}, which uses both facilities so that the citation
appears with three authors, and ``Sepublica 2012'' as the container title.

Since, kblog-metadata was released, WordPress also supports ``Guest authors''
through the co-authors-plus plugin -- which likewise dissociates login rights
from authorship; this provides a much nicer graphical environment for defining
co-authors than kblog-metadata, but comes with an overhead that authors must
be created individually. Kblog-metadata will use metadata from this plugin if
it is installed.

Finally, we have added support for the use of shortcodes to define authorship.
This is very useful when content is being generated outside of the WordPress
environment; for example, on \url{http://bio-ontologies.knowledgeblog.org},
most of the content is generated from Word documents. During publication, we
markup the author names with shortcodes --- \texttt{[author]Phillip
  Lord[/author]}; this markup passes unmolested through Word's HTML conversion
and is then interpreted by WordPress. This prevents cut-and-paste errors that
would occur if authors had to be added manually --- a significant issue for
science where most articles have many authors. This website also modifies the
container title to distinguish between different years.

\section{Identifying by Proxy}
\label{sec:referencing-proxy}

One significant issue with kcite as a referencing engine is the requirement
for a primary identifier for every item\footnote{kcite does allow addition of
  all citation metadata within an inline shortcode, although this is intended
  as a fallback}. Most scientific literature, and any article posted on the
web is likely to have an identifier that kcite can use. However, this causes
problems for two specific types of resource. First, many smaller conferences
and workshops do not publish their literature in a web capable form; in many
cases papers on the web are available as PDF or Postscript only. And even when
web hosted, sites may not add bibliographic metadata. Kblog-metadata provides
a partial solution to these problems: authors can host their articles, and
alter the metadata accordingly as described in
Section~\ref{sec:kblog-metadata}. However, this fails for work by other
authors, whose work cannot be posted without permission. A similar problem
exists for books; while these generally do have a standard identifier (ISBN)
we have not been able to find a publicly available mechanism to automate the
transformation from ISCN to structured bibliographic metadata.

Greycite provides a mechanism to address this difficulty. There are a number
of catalogues available for both scientific literature and books; these often
have a primary URI which can be used as a reference identifier. Greycite
currently supports several sites of this form: WorldCat
(\url{http://worldcat.org}) provides URIs for books (as well as other forms of
media such as CDs and DVDs), Mendeley which references journal articles and
OpenLibrary (\url{http://www.openlibrary.org}) which also provides URIs for
books. In these cases, references will appear correctly when used in Kcite,
showing the source of metadata which could, in principle, be used to track the
original resource.

\section{Metadata for Preservation}
\label{sec:preservation}

One recurrent issue with author-led publishing is the difficulties associated
with digital preservation; custom and practice means that it is considerably
harder for author-led publications to ensure that work is preserved than
third-party publications; systems such as CLOCKKS or LOCKKS are often just not
accessible to smaller-scale author-led publication.

To address this need, we have integrated greycite with public archiving
efforts, such as the Internet Archive, the UK Web Archive and WebCite. As well
as scanning URIs for metadata, greycite periodically checks these archive
sites, to see if they are available as archives. We use this metadata in a
number of ways. 

First, archive sites are available directly from Greycite through a REST API.
Kblog-metadata provides an ``archived'' widget where it publicly displays
this information; this provides a third-party stamp that the article has
actually been available from the time stated, as well as an \textit{ad hoc}
enforcement of non-repudiability. If an author changes their own content, the
differences with the archived sites will be clear. 

If a site disappears, then these links to the archives will also disappear.
Kblog-metadata also allows readers to download BibTeX files for any (or all)
articles; this metadata comes directly from Greycite and includes links to all
known public archives. Anyone citing an article using this file will therefore
have a reference to archival versions.

Of the services we currently check for archival versions, currently only
WebCite offers on-demand archiving\footnote{WebCite is asking for funding on
  the web, which is an unfortunate sign}. Greycite currently submits any
archive with four (TCDA) piece of metadata to WebCite for archiving.
Additionally, greycite itself stores historical metadata for indeterminate
amounts of time, and therefore constitutes a metadata archive.

\section{Tracking movement around the web}
\label{sec:tracking}

In addition to the four pieces of bibliographic metadata, greycite collects
one other key piece of knowledge; a canonical URI. Currently, this knowledge
is not represented in many of the formats we harvest. While, CoINS does
provide a field which can be used for this purpose, in practice it is not that
useful: CoINS is used to embed bibliographic metadata into the web, but the
CoINS may not relate to the article in which it is embedded. Open Graph
Protocol data also returns an explicit identifier; in this case, this is about
the article in question. This means Greycite can store a canonical URI for a
particular article, regardless of the URI used to access the article. Again,
and perhaps unexpectedly, RSS/Atom feeds are extremely useful; these carry a
link and explicitly state whether it is a permalink (i.e. canonical) or not.

The presence of a canonical URI makes it possible to track content as it
moves. For instance, it is relatively common for blogs to change their
permalink structure; with WordPress, for instance, existing links are
maintained through the use of a 301 Redirect response. Greycite could
recognise this situation and use the Redirect location as the canonical link;
unfortunately HTTP redirects are used for many different purposes, including
load balancing. Instead, greycite recognises that the URI used to fetch a
request and the stated canonical URI are different and records this fact. For
example, greycite records that
\url{http://www.russet.org.uk/blog/2012/02/kcite-spreads-its-wings/} changed
from being canonical to not sometime between April 2012 and Jan 2013 (actually
this happened in June 2012).

Currently, greycite returns the canonical URI with requests for both BibTeX or
JSON data; authors of referring documents will therefore will have an recent
link. Although, currently not implemented, we plan to add more explicit
support for this to our kcite client, so that it will display canonical URIs;
again, this supports digital preservation. Articles which refer to URIs which
have ceased to be canonical, would be able to display both the URI
to which the author original referred and the correct canonical reference. 

The ability to track articles as they move also opens up a second possibility.
Currently, one main stated advantage of systems like DOIs is the ability to
change the location of a record without necessitating a change in identifier.
A similar system is also available in the form of PURLs (persistent
URLs)\footnote{\url{http://purl.oclc.org/}}. Greycite allocates PURLs for all
URIs for which it can extract all the required piece of metadata. Currently,
these redirect to the last known canonical URI for a given URI; in effect,
this means that PURLs will track URIs for any website that maintains its
redirects and metadata for sufficient time for Greycite to discover this.

\section{Discussion}
\label{sec:discussion}

As we have previously stated\cite{greycite1325}, our belief is that semantic
metadata, if it is to be useful at all, must be useful to all the key players
in the publication process; critically, this includes the author. The tools
that we have described here obey this doctrine; we seek to aid and reward the
authors who use good metadata. 

Kcite already follows this principle: if links are inaccurate, then the
reference will not format correctly (or at all). As well as errors made during
authoring, we (PWL) have found non-functioning DOIs, as well as one systematic
error in DOI presentation which has resulted in a change to CrossRef display
guidelines\cite{greycite8865}. In addition, correct formatting of the
references depends on the metadata being correct. Again, here, we have found
DOIs with inaccurate metadata. With the addition of greycite, this
functionality has been extended to any URI. Authors are very likely to cite
themselves. If they do so, they are now dependent on their own metadata; if
the metadata is wrong, then references will be. This provides an
incentive for authors to correct metadata for their own purposes,
simultaneously making everyone's life better\footnote{slightly}. As well as
correcting our own websites, use of greycite has discovered inaccurate
metadata in commercial publishing websites.

Greycite is currently unique so suffers from some of the limitations of
centralisation; however, effectively, it is just a cache. The metadata that it
provides is sourced from the distributed resources that are referenced; it can
support multiple installations trivially. Except in the case of ephemeral
metadata, none of these would be privileged. The current implementation of
greycite also provides an initial answer to our question, is there enough
bibliographic metadata on the web to enable citation: a qualified yes. Through
the use of existing metadata schemes and some heuristics, we can discover this
metadata for many websites. An early analysis suggests that greycite can
provide the four key pieces of metadata for around 1\% of the web, which
consitutes 100s of millions of URIs; the percentage for ``interesting''
websites is much higher, at over 60\%. We currently also lack any statistical
analysis on how correct this metadata is; by inspection, the level of
correctness within the $\sim$4000 URIs submitted from 254 independent IP
addresses is high, but this result is biased as we have corrected errors
iteratively. For random URIs, we lack a gold standard, and most are not in
English making inspection hard.

While using metadata to generate references is useful, it is one-step removed.
The author is not supported in discovering that their metadata is inaccurate
until sometime after it has been published. Kblog-metadata now improves on
this process and makes it more immediate; by visualising the metadata on
publication, authors can check that it is correct. Likewise, the same metadata
is used to generate a BibTeX file which they can use. As an open source tool,
it is hard to know how many installations kblog-metadata currently has,
although download statistics would suggest 30 or 40, including one journal.

Set against this desire to improve the quality of metadata on the web,
greycite has taken a pragmatic approach to the metadata standards it uses. It
currently supports many of the different ways of marking up bibliographic
metadata. More over, it uses many heuristics, to cope with metadata which is
unclean or just broken. This works against the notion of encouraging authors
to improve their metadata; however, increasing the utility of the API makes
this a compromise well worth making.

We are also addressing the issue of digital preservation; we achieve this in
two ways. First, we leverage existing web archives, deep linking through to
them where content has already been archived. To achieve this in a simple
manner requires no semantics at all, beyond the URI for a given resource.
However, resources may be present at more than one URI, or may change their
canonical URI over time. Greycite is now making preliminary use of this
metadata to track articles as they move; the current location can be retrieved
by a client, or alternatively greycite provides PURLs which will work with any
client. 

As with our previous work, the level of semantics provided or used by these
publication tools is not high; however, by using existing metadata standards,
greycite can provide metadata for 100s of millions of URIs including many from
websites which are unlikely to care about academic referencing. We have
focused on adding value for authors, both when referencing or displaying
citations on an article. By adding value for the authors, we help to ensure
that they will add value to the metadata. While this approach adds very small
amounts of metadata for an individual article, the aggregate total of metadata
over all articles is, potentially, vast.

\renewcommand{\UrlFont}{\footnotesize}
\bibliographystyle{splncs}
\bibliography{phil_lord_refs,phil_lord_all,russet,urls}

\end{document}